# Applications of generating function method to the symmetry and the Kronecker products of SU(n) representations.

M. Hage-Hassan
Université Libanaise, Faculté des Sciences Section (1)
Hadath-Beyrouth

## Abstract

Using the generating function of SU(n) we find the conjugate state of SU(n) basis and we find in terms of Gel'fand basis of SU(3(n-1)) the representation of the invariants of the kronecker products of SU(n). We find a formula for the number of the elementary invariants of SU(n). We apply our method to the coupling of SU(3) and we find a new expression of the isoscalar of Wigner symbols ($\lambda_1 0,\lambda_2\mu_2;\ \lambda_3\mu_3$).

## 1. Introduction

The applications of the SU (n) group theory in nuclear physics have occurred in particles and for description of nuclear collective properties and for the classification of elementary particles[1-3]. These applications require the determination of the representations and the Clebsh-Gordan coefficients, or Wigner symbols, of unitary groups [3]. The bases of SU (2) and SU (3) were determined by Cartan-Weyl method [3]. And Gel'fand and Zeitlin [4-6], using the Young diagram of Schur-Weyl, found an orthogonal basis of SU (n) consisting of positive numbers $n(n+1)/2$ :

$$\left|(h)_n\right\rangle = \left|\begin{matrix}[h]_n \\ (h)_{n-1}\end{matrix}\right\rangle \text{With } [h]_n = [h_{1n}, .., h_{nn}]_n. \tag{1,1}$$

But it is important to note that E. Cartan [5] already proved that any arbitrary irreducible representation of U (n) can be expressed in terms of vectors ${}_n\Delta_i^m$ of a set of n-subspaces $\{{}_n\Delta^m\}$ called the fundamental representations:

$$\{{}_n\Delta^1\} = [100...0]_n,\ \{{}_n\Delta^2\} = [110...0]_n, ..., \{{}_n\Delta^n\} = [111,...,11]_n. \tag{1,2}$$

We observed that the vectors $\{{}_n\Delta_i^m\}$ may be expressed in terms of binary numbers with the sum of its one is m and we called it by the binary fundamentals basis (BFR) [7]. Generalizing the generating functions of the oscillator, SU (2), SU (3) we simply find, with the help of the BFR, the generating function of Gel'fand basis of SU (n) [7-8].

In this paper we write the generating function in terms of the binary basis then by applying the transformation of binary numbers by its complement in this function we find simply the conjugate [6] and sub conjugate states of SU (n).

Several approaches have been proposed for the calculation of Wigner symbols: The infinitesimal approach [3], the Young diagram of Schur-Weyl [3] and the tensor operator method has been developed extensively by several research groups[9-14].

Finally Van der Wearden has shown that the Wigner symbols can be determine from the development of invariants functions{H}formed from the elementary invariants of the kronecker product of three representations of SU (2)[15-16].
The application of this method to SU (3) has been studied by several authors [17-18]. But Moshinsky observed that the kronecker product of the representations of SU (n) could be analyzed in terms of some representations of SU (N) where N =K(n-1) [19-21].

Taking this observation into consideration and using the generating function in the case of SU (3) multiplicity free we obtained an expression of Wigner symbols with a single sum instead of two summations by the tensor operator method [22].More {H} are orthogonal functions and we find it as a vector of the representation of SU(3).
This important results has encouraged us to develop and extend our generating function method for the calculation of Wigner symbols of SU (n) [23]. We find in this paper the invariant functions among the vectors of Gel'fand basis of SU (3(n-1)) and a formula for the number of invariants. We also found a new expression of isoscalar factors of the tensor product of SU (3) with two summations.

In this paper we summarize the basis of SU(2),SU(3), the base of Gel'fand and generator of SU(n) in parts two, three and four. In part five we determine the Gel'fand conjugate state of SU(n). We study the Wigner of symbols in part six. In the seven and eight parts we give the bases of Gel'fand invariants of SU(2) and SU(3). The basis of Gel'fand invariants of SU (n) and the formula for the number of invariants will be given in part nine. In part ten we expose the derivation of the isoscalar of SU(3).

## 2. The group SU (2) ⊂ SU (3)

We will give only a summary of the basis of SU (2) and SU (3).

### *2.1 The basis of the group SU (2) in the analytic Hilbert space*

It's well known [5] that the basis of SU(2) is

$$\varphi_{jm}(z) = \frac{z_1^{j+m} z_2^{j-m}}{\sqrt{(j+m)!(j-m)!}} \qquad (2,1)$$

$\varphi_{jm}(z)$ is the basis of the analytic Hilbert space, the Fock or the Fock-Bargmann(F-B) space, with the Gaussian measure:

$$d\mu(z) = \exp(-z \cdot \bar{z}) dz_1 dz_2, \ dz_i = dx_i dy_i, \ z_i = x_i + iy_i = {}_2\Delta_i^1 \qquad (2,2)$$

### *2.2 The basis of the group SU (2) ⊂ SU (3)*

Let $D_{[\lambda,\mu]}$ the space of homogeneous polynomials and $V^{\lambda\mu}_{(t,tz,y)}(z^1, z^2)$ is the orthogonal basis with:

$$z^{(i)} = (z_1^i, z_2^i, z_3^i), \ z_{(j)} = (z_j^1, z_j^2, z_j^3), \ z_i^j \in C$$

$$V^{\lambda\mu}_{(t,tz,y)}(z^{(1)}, z^{(2)}) = V^{\lambda\mu}_{(t,tz,y)}(z_{(1)}, z_{(2)}, z_{(3)})$$

And $T_{ij} = \sum_k z^i_k (\partial / \partial z^i_k)$ , $T^{ij} = \sum_k z^k_i (\partial / \partial z^k_j)$ (2,3)

A- $V^{\lambda\mu}_{(t,tz,y)}(z^1, z^2)$ is homogeneous with respect to $z^{(i)}$ then

$$T_{11} V^{\lambda\mu}_{(\alpha)} = (\lambda + \mu) V^{\lambda\mu}_{(\alpha)}, \qquad T_{22} V^{\lambda\mu}_{(\alpha)} = (\mu) V^{\lambda\mu}_{(\alpha)} \tag{2,4}$$

The vectors $V^{\lambda\mu}_{(t,tz,y)}(z^1, z^2)$ are eigenfunctions of the Casimir operator of the second order $\vec{T}^2$, the projection of $\vec{T}$ on the z axis and the hypercharge Y. The eigenvalue of these operators are respectively t (t + 1), $t_z$ and the triple of the hypercharge quantum number y. The numbers $t, t_z$ are the isospin and the component of isospin on the z axis.
We have:

$$Y V^{\lambda\mu}_{(\alpha)} = y V^{\lambda\mu}_{(\alpha)}, \qquad T_z V^{\lambda\mu}_{(\alpha)} = t_z V^{\lambda\mu}_{(\alpha)} \tag{2,5}$$

And

$$\vec{T}^2 V^{\lambda\mu}_{(\alpha)} = t(t+1) V^{\lambda\mu}_{(\alpha)}. \tag{2,6}$$

B- The expression of $V^{\lambda\mu}_{(\alpha)}(z^1, z^2)$ is well known [1- 19] and we give only the result:

$$V^{\lambda\mu}_{(t,tz,y)}(z^1, z^2) = N(\lambda\mu;\alpha)(-1)^s \times \sum_k \binom{m}{k} \frac{(\mu - s)!r!}{(\mu - s - k)![r - (n-k)]!}$$
$$\times ({}_3\Delta^1_1)^{r-(m-k)} ({}_3\Delta^1_2)^{m-k} ({}_3\Delta^1_3)^{\lambda - r} ({}_3\Delta^2_{23})^k ({}_3\Delta^2_{13})^{\mu - s - k} ({}_3\Delta^2_{12})^s \tag{2,7}$$

Put $(z^1) = ({}_3\Delta^1_1, {}_3\Delta^1_2, {}_3\Delta^1_3)$, $z^1 \times z^2 = ({}_3\Delta^2_{23}, -{}_3\Delta^2_{13}, {}_3\Delta^2_{12})$

We write: $$V^{\lambda\mu}_{(t,tz,y)}({}_3\Delta^1, {}_3\Delta^{12}) = V^{\lambda\mu}_{(t,tz,y)}(z^1, z^2) \tag{2,8}$$

And $$N(\lambda\mu;\alpha) = \left\{ \frac{(\lambda+1)!(\mu + r - s + 1)!}{r!s!(\mu - s)!(\lambda - r)!(\mu + r + 1)!(\lambda + \mu - s + 1)!} \times \frac{(2t - m)!}{(2t)!m!} \right\}^{\frac{1}{2}}$$

With $y = -(2\lambda + \mu) + 3(r + s),\ 0 \le r \le \lambda,$

$2t = \mu + (r - s),\quad 0 \le s \le \mu,$

And $t_z = t - m,\quad m = 0, 1, \ldots, 2t.$ (2,9)

C-$V^{\lambda\mu}_{(t,tz,y)}({}_3\Delta^1, {}_3\Delta^2)$ is also homogeneous with respect to $z_{(i)}$ then:

$$T^{11} V^{\lambda\mu}_{(t,tz,y)} = \left(t_z + \frac{y + (\lambda + 2\mu)}{3}\right) V^{\lambda\mu}_{(t,tz,y)},$$
$$T^{22} V^{\lambda\mu}_{(t,tz,y)} = \left(-t_z + \frac{y + (\lambda + 2\mu)}{3}\right) V^{\lambda\mu}_{(t,tz,y)},$$
$$T^{33} V^{\lambda\mu}_{(t,tz,y)}(z^1, z^2) = \frac{-y + (\lambda + 2\mu)}{3} V^{\lambda\mu}_{(t,tz,y)}(z^1, z^2) \tag{2,10}$$

# 3. The Gel'fand basis of unitary group

***3.1 The Gel'fand basis of the unitary groups***

Using the " Weyl's branching law" Gelfand-Zeitlin [4-6] introduced the basis of representation of U (n), function of $n(n+1)/2$ integers numbers.

$$\left|(h)_n\right\rangle = \left|\begin{matrix} h_{1n} & h_{2n} & \cdots & h_{nn} \\ & h_{1n-1} & \cdots\cdots & \\ & h_{12} & h_{22} & \\ & & h_{11} & \end{matrix}\right\rangle = \left|\begin{matrix} [h]_n \\ (h)_{n-1} \end{matrix}\right\rangle \qquad (3,1)$$

With $h_{i,j} \geq h_{i,j-1} \geq h_{i+1,j},\ j = 2,...,n\ \ et\ \ i = 1,..., j.$

And $h_{nn}=0$ for SU(n)

***3.2 Cartan fundamental representations and the binary numbers***

**3.2.*1 The fundamental representations:***

The fundamental representations $\{\Delta_n^m\}$ of U (n) are the irreducible subspace $[h]_n$:

$$[\overbrace{1,1,1,..,1}^{m},0,...,0]_n = {}_n\Delta^m \qquad (3,2)$$

${}_n\Delta_i^m$ , $i = 1,...,C_n^m$ Are the vectors of the subspace $\{{}_n\Delta^m\}$ .

We give only the correspondence between the binary representation and the variables ${}_4\Delta_i^m$ of SU(4) as an example:

| | | | | |
|---|---|---|---|---|
| ${}_4\Delta_i^1$ | 1 0 0 0 | 0 1 0 0 | 0 0 1 0 | 0 0 0 1 |
| ${}_4\Delta_i^2$ | 1 1 0 0 | 1 0 1 0 | 1 0 0 1 | |
| ${}_4\Delta_i^2$ | 0 0 1 1 | 0 1 0 1 | 0 1 1 0 | |
| ${}_4\Delta_i^3$ | 0 1 1 1 | 1 0 1 1 | 1 1 0 1 | 1 1 1 0 |

***Table 1***

***3.2.2 The representations of BFR in the analytic Hilbert space***

We will generalize the basis vectors and the binary basis of SU(2) and SU(3):

| SU(2) Basis ${}_2\Delta_i^1$ | 1-$\Delta_1^1$ | 2-$\Delta_2^1$ |
|---|---|---|
| SU(2) Binary basis | 1 0 | 0 1 |

| SU(3) basis ${}_3\Delta_i^m$ | 1-$\Delta_1^1$ | 2-$\Delta_2^1$ | 3-$\Delta_3^1$ | 4-$\Delta_{23}^{12}$ | 5-$\Delta_{13}^{12}$ | 6-$\Delta_{12}^{12}$ |
|---|---|---|---|---|---|---|
| SU(3) Binary basis | 1 0 0 | 0 1 0 | 0 0 1 | 0 1 1 | 1 0 1 | 1 1 0 |

***Table2***

Let $(z_i^j),\ i, j = 1,...,n$ a matrix of complexes numbers and we consider the minors of this matrix: We associate to each miner $\Delta_{i_1...i_l}^{12...l}$ of the matrix $(z_i^j)$ i, j=1,..,na table of n-boxes numbered from 1 to n. We put "one" in the boxes $i_1, i_2, \ldots, i_l$ and zeros elsewhere.

$$
\begin{array}{ccccccccc} & 1 & 2 & \dots & i_1 & \dots & i_k & \dots & n \end{array}
$$

$$
{}_n\Delta_l^k(z) = \Delta_{i_1\dots i_k}^{12\dots k} = \left| \begin{array}{cccccccc} 0 & 0 & \dots & 1 & \dots & 1 & \dots & 0 \end{array} \right| (z)
$$

The supplement is: $${}_n\overline{\Delta}_i^k(z) = \overline{\Delta}_{i_1\dots i_k}^{12\dots k} = \left| \begin{array}{cccccccc} 1 & 1 & \dots & 0 & \dots & 0 & \dots & 1 \end{array} \right| (z) \qquad (3,3)$$

The polynomials $\left\{ {}_n\Delta_i^k(z) \right\}$ form an orthogonal basis and it follows that the vectors of BFR are orthogonal.

The Gel'fand representation can be written in the analytic Hilbert or Fock space by:

$$
\Gamma_n(\Delta(z)) = \langle (h)_n \| \Delta(z) \rangle, \quad and \quad \Gamma_3(\Delta(z)) = V_{(t,tz,y)}^{\lambda\mu}({}_3\Delta^1, {}_3\Delta^2) \qquad (3,4)
$$

## 4. The generating function of Gel'fand basis

We review the generating functions of SU (2), SU (3) and we determine the generating function of Gel'fand basis.

### *4.1 Generating function of SU(2) and the Gel'fand basis*

We express the generating function of SU(2) in terms of Gel'fand indices by:

$$
G(\xi, z) = \sum_{h_{11}, h_{12}} \frac{(x_2^1)^{h_{12}-h_{11}} (y_2^1)^{h_{11}}}{\sqrt{(h_{12}-h_{11})!(h_{11})!}} \Gamma_2 \begin{pmatrix} h_{12} & & 0 \\ & h_{11} & \end{pmatrix} (\Delta(z)) = \exp[y_2^1 \; {}_2\Delta_1^1 + x_2^1 \; {}_2\Delta_2^1] \qquad (4,1)
$$

With: $\quad h_{12} = 2j, \quad h_{11} = j - m, \quad h_{22} = 0$.

### *4.2 The generating function of SU(3)*

The generating function [7-8] of SU (3) may be written in Fock-Bargmann basis by:

$$
\sum_{\lambda\mu} \prod_{\lambda=2}^{3} \prod_{\mu=1}^{\lambda-1} A_{\lambda\mu}((x_\lambda^\mu)^{h_{\mu\lambda}-h_{\mu\lambda-1}} (y_\lambda^\mu)^{h_{\mu\lambda-1}-h_{\mu+1\lambda}}) \, \Gamma_3 \begin{pmatrix} [h]_3 \\ (h)_3 \end{pmatrix} (\Delta(z) =
$$

$$
\exp[{}_3\Delta_{12}^{(12)} y_3^2 + z_3^2({}_3\Delta_{23}^{(12)} x_2^1 + {}_3\Delta_{13}^{(12)} y_2^1) + y_3^1({}_3\Delta_1^1 y_2^1 + {}_3\Delta_2^1 x_2^1) + {}_3\Delta_3^1 x_3^1] \qquad (4,2)
$$

We observed [8] that the powers of parameters x and y have the same powers of raising and lowering operators introduced by Nagel and Moshinsky [20].
In Gel'fand notations [7] we write (2,6) as:

$$
\begin{pmatrix} [h]_3 \\ (h)_2 \end{pmatrix} = \begin{pmatrix} \lambda + \mu & & \mu & & 0 \\ & t + \frac{Y}{2} + \frac{(\lambda + 2\mu)}{3} & & -t + \frac{Y}{2} + \frac{(\lambda + 2\mu)}{3} & \\ & & t_z + \frac{Y}{2} + \frac{(\lambda + 2\mu)}{3} & & \end{pmatrix} \qquad (4,3)
$$

### *4.3 The generating function of SU(n)*

Generalizing the generating functions of the oscillator, SU (2) and SU (3) we write:

$$\sum_{\lambda\mu} \prod_{\lambda=2}^{n} \prod_{\mu=1}^{\lambda-1} A_{\lambda\mu} \left( (x_{\lambda}^{\mu})^{h_{\mu\,\lambda}-h_{\mu\,\lambda-1}} (y_{\lambda}^{\mu})^{h_{\mu\,\lambda-1}-h_{\mu+1\,\lambda}} \right) \Gamma_n(\Delta(z))$$

$$= \exp[\ \sum_{m,i} \varphi_{n,i}^{m}(x,y)\, {}_n\Delta_j^m(z)], \qquad (4,4)$$

With $\quad e(\mu,\lambda) = h_{\mu\,\lambda} - h_{\mu\,\lambda-1},\ \ and\ \ f(\mu,\lambda) = h_{\mu\lambda-1} - h_{\mu+1\,\lambda}$.

### *4.4 Calculus of the coefficients* $\varphi_{n,i}^{m}(x,y)$

The coefficients $\varphi_{n,i}^{m}(x,y)$ may be written as product of parameters $y_{\lambda}^{\mu} = y(\lambda,\mu)$ and $x_{\lambda}^{\mu} = x(\lambda,\mu)$. We determine the indices of these parameters by using the following rules:

a- We associate to each "one" which appeared after the first zero a parameter $x_{\lambda}^{\mu}$ whose index $\lambda$ is the number of boxes and $\mu$ the number of "one" before him, plus one.

b- We associate to each zero after the first "one" a parameter $y_{\lambda}^{\mu}$ whose index $\lambda$ is the number of boxes and $\mu$ is the number of "one" before him.

### *4.5 The coefficients* $\varphi_{n,i}^{m}(x,y)$ *of SU(2) and SU(3)*

| SU(2) Binary basis | 1 0 | 0 1 |
|---|---|---|
| $\varphi_{2,I}$ | $y_2^1$ | $x_2^1$ |

| SU(3) Binary basis | 1 0 0 | 0 1 0 | 0 0 1 | 0 1 1 | 1 0 1 | 1 1 0 |
|---|---|---|---|---|---|---|
| $\varphi_{3,I}$ | $y_2^1 y_3^1$ | $x_2^1 y_3^1$ | $x_3^1$ | $x_2^1 x_3^2$ | $y_2^1 x_3^2$ | $y_3^2$ |

***Table 3***

## 5. The symmetry and the conjugate states of SU(n)

We know that each binary number has a complement then we deduce that $\Delta_{n,[i]}^{k}$ has a complement $\overline{\Delta}_{n,[i]}^{k}$, Therefore the generating function is invariant by the transformation:

$${}_n\Delta_i^k \rightarrow {}_n\overline{\Delta}_i^k \qquad (5,1)$$

### *5.1 The conjugate state* $\Gamma_n\left((h^c)_n\right)$ *of Gel'fand basis*

The generating function of the conjugate state $\Gamma_n\left((h^c)_n\right)$ becomes:

$$\sum_{\lambda\mu} \prod_{\lambda=2}^{n} \prod_{\mu=1}^{\lambda-1} A_{\lambda\mu} \left( (x_{\lambda}^{\mu})^{h_{\mu\,\lambda}-h_{\mu\,\lambda-1}} (y_{\lambda}^{\mu})^{h_{\mu\,\lambda-1}-h_{\mu+1\,\lambda}} \right) \Gamma_n\left((h)_n\right)_c$$

$$= \exp\left[\ \sum_{m,i} \varphi_{n,i}^{m}(x,y)\, {}_n\overline{\Delta}_j^m\ \right] \qquad (5,2)$$

We change $x_\lambda^\mu$ *and* $y_\lambda^\mu$ *by* $y_\lambda^{\lambda-\mu}$ *and* $x_\lambda^{\lambda-\mu}$ in $\varphi_{n,i}^m(x,y)$ to get $\varphi'^m_{n,i}(x,y)$.
Put ν=λ-μ so we write:

$$\sum_{\lambda\mu} \prod_{\lambda=2}^{n} \prod_{\mu=1}^{\lambda-1} A_{\lambda,\nu}\left((x_\lambda^\nu)^{h^c_{\nu\,\lambda}-h^c_{\nu\,\lambda-1}}(y_\lambda^\nu)^{h^c_{\nu\,\lambda-1}-h^c_{\nu+1\,\lambda}}\right)\Gamma_n\left((h^c)_n\right)$$

$$= \exp[\ \sum_{m,i} \varphi'^m_{n,i}\ (x,y) \times_n \overline{\Delta}^m_j\ ], \tag{5,3}$$

Comparing (5,2) and (5,3) we find:

$$\boxed{h^c_{\mu,\lambda-1} - h^c_{\mu+1,\lambda} = h_{\lambda-\mu,\lambda} - h_{\lambda-\mu,\lambda-1}}\ \boxed{h^c_{\mu,\lambda} - h^c_{\mu,\lambda-1} = h_{\lambda-\mu,\lambda-1} - h_{\lambda-\mu+1,\lambda}},$$

$$\boxed{h^c_{n,n} = h_{n,n} = 0} \tag{5,4}$$

We will determine the conjugate state $\left|(h^c)_n\right\rangle$ with the help of (5,4).

***5.2 Expressions of the indices of the conjugate states***:

We proceed by induction to determine the indices of the conjugate states:

Let λ=n

μ=n-1 $h^c_{n-1,n-1} - h^c_{n,n} = h_{1,n} - h_{1,n-1}$ Then $\boxed{h^c_{n-1,n-1} = h_{1,n} - h_{1,n-1}}$

$h^c_{n-1,n} - h^c_{n-1,n-1} = h_{1,n-1} - h_{2,n}$ Then $\boxed{h^c_{n-1,n} = h_{1,n} - h_{2,n}}$

For μ=n-2 $h^c_{n-2,n-1} = h_{1,n} - h_{2,n-1}$ , $h^c_{n-2,n} = h_{1,n} - h_{3,n}$,

…………………………………………

After the calculations for λ= n, n-1 .., 1 and μ= 1,..,λ-1 we find:

$$\boxed{h^c_{i,j} = h_{1,n} - h_{j-i+1,j}} \tag{5,5}$$

***5.3 The phase factor***

By extension of phase factor φ of SU (2) [6], we write:

$$\boxed{\varphi = \sum_{i,j} h_{i,j} - h_{1n}} \tag{5,6}$$

***-The conjugate state of SU (2) and SU(3)***

- The conjugate state of SU (2) is:

$$h^c_{12} = 2j,\ h^c_{22} = 0,\ h^c_{11} = j+m$$

- The conjugate state of SU (3) is the transformation : (t, $t_z$, Y) → (t, -$t_z$, -Y)

So we get the conjugate basis and the R-Conjugation of Gell-Mann of SU (3) [17].

***5.4 The sub conjugate state of SU (n-1) ⊂ SU (n)***

For this case we retain the binary number n and we take only the complement of the first n-1 cells. We find that this transformation requires the following conditions:

$$e^c(1,n) = e(1,n), f^c(n-1,n) = f(n-1,n)$$
$$e^c(2,n) = e(n-1,n),\ e^c(n-1,n) = e(2,n),$$
$$e^c(n-i-j,\ n-i) = f(j,n-i),\ f^c(j,n-i) = e(n-i-j,\ n-i) \tag{5,7}$$

With $i = 1,..,(n-2),\ j < i$.

We find for SU (3) the transformation (t, $t_z$, Y) ->(t, -$t_z$, Y).

***5.5 The conjugate states and unitary the transformation***

We denote the unitary representation matrix of the unitary transformation U by:

$$D^{[h]_n}_{(h'),(h)}(U) = \left\langle \begin{matrix}[h]_n\\(h')_n\end{matrix} \middle| U \middle| \begin{matrix}[h]_n\\(h)_n\end{matrix} \right\rangle \tag{5,8}$$

We have :

$$U\left|\begin{matrix}[h]_n\\(h)_n\end{matrix}\right\rangle = \sum_{h'} D^{[h]_n}_{(h'),(h)}(U)\left|\begin{matrix}[h]_n\\(h')_n\end{matrix}\right\rangle \quad \text{And} \quad U\left|\begin{matrix}[h]_n\\(h)_n\end{matrix}\right\rangle_c = \sum_{h'} \overline{D}^{[h]_n}_{(h'),(h)}(U)\left|\begin{matrix}[h]_n\\(h')_n\end{matrix}\right\rangle_c \tag{5,9}$$

With $\overline{D}^{[h]_n}_{(h'),(h)}(U) = D^{[h]_n}_{(h'),(h)}(U^*)$ is the complex conjugates of $D^{[h]_n}_{(h'),(h)}(U)$.

# 6. Wigner's symbols and the invariants of SU(n)

In this section we give the definition of invariants and its connection with the Wigner coefficients [21-24]. By using the binary representation of invariants and the parameter space we show that our method gives the Van der Wearden's result of SU(2).

***6.1 Wigner's symbols and the generalization of Van der Wearden's invariants***

***6.1.1 Wigner's symbols***

The direct product of two representations may be reduced according to the formula

$$[h^1]\otimes[h^2] = \sum(\rho)[h^3]_{(\rho)} \tag{6,1}$$

Where (ρ) is the multiplicity or the number of time the representation is contained in $[h^1]\otimes[h^2]$. With

$$\left|\begin{matrix}[h^3]\\(h^3)\end{matrix}\right\rangle_\rho = \sum_{h^1h^2}\left\langle\begin{matrix}[h^1] & [h^2]\\(h^1) & (h^2)\end{matrix}\middle\|\begin{matrix}[h^3]\\(h^3)\end{matrix}\right\rangle_\rho \times \left|\begin{matrix}[h^1]\\(h^1)\end{matrix}\right\rangle\left|\begin{matrix}[h^2]\\(h^2)\end{matrix}\right\rangle \tag{6,2}$$

The coefficients in this expression are the Clebsh-Gordan coefficients.

***6.1.2 The invariant of the unitary transformation***

The vector
$$\frac{1}{\sqrt{d_{h^3}}}\sum_{(h^3)}\left|\begin{matrix}[h^3]\\(h^3)\end{matrix}\right\rangle_\rho\left|\begin{matrix}[h^3]\\(h^3)\end{matrix}\right\rangle_c \tag{6,3}$$

is an invariant by unitary transformation with unity norm in the product of three spaces. When we replace it with the equation (6,2) we write:

$$H_{(\rho)}=\sum_{h^1h^2}\begin{pmatrix}[h^1] & [h^2] & [h^3]\\(h^1) & (h^2) & (h^3)\end{pmatrix}_\rho\left(\prod_{i=1}^{2}\Gamma_n\begin{pmatrix}[h^i]\\(h^i)\end{pmatrix}\left({}^s\Delta_n^i(z)\right)\right)\Gamma_n\begin{pmatrix}[h^3]\\(h^3)\end{pmatrix}\left({}^s\Delta_n^3(z)\right)_c \qquad (6,4)$$

The invariant polynomials $H_{(\rho)}$ is a function of the elementary invariants $\left({}^s\Delta_n^i(z)\right)$.

The coefficients $$\begin{pmatrix}[h^1] & [h^2] & [h^3]\\(h^1) & (h^2) & (h^3)\end{pmatrix}_\rho=\frac{1}{\sqrt{d_{h^3}}}\left\langle\begin{matrix}[h^1] & [h^2]\\(h^1) & (h^2)\end{matrix}\middle\|\begin{matrix}[h^3]\\(h^3)\end{matrix}\right\rangle_{c\rho} \qquad (6,5)$$

Are Wigner's 3j symbols of SU (n) and ρ is the indices of multiplicity.
$H_{(\rho)}$ is the generalization of the Van der Wearden's invariant of the group SU(2).
Using (6,4) and (6,3) we find the following properties:

$$U^{(1,2,3)}H_{(\rho)}=H_{(\rho)},\text{And}\quad \left\langle H_{(\rho)}\middle|H_{(\rho')}\right\rangle=\delta_{(\rho),(\rho')} \qquad (6,6)$$

Then we can choose $H_{(\rho)}$ as subspace of SU(3 (n-1)) which are function of the compatible elementary invariants ${}^s\Delta_n^i(z)$.

### *6.2 The elementary invariants ${}^s\Delta_n^i(z)$ and ${}^s\phi_n^i$*

We determine the elementary scalars ${}^s\Delta_n^i(z)$ which are the basic elements of the Gel'fand basis of the SU (k (n-1)). These scalars are formed of k rows of tables, Where each row of (n-1) boxes and $\alpha_i$ "one" and zero elsewhere.
$\alpha_i$ Satisfies the following conditions

$$0\le\alpha_i\le n-1,\ \sum_{i=1}^{k}\alpha_i=n \qquad (6,7)$$

## *7. The Gel'fand basis for the invariants of SU(2)*

We will give all intermediate calculations to clarify the generalization to SU(n).

### *7.1 The elementary invariants*

We find for SU (2) the three elementary invariants in the Gel'fand basis

$$\overline{|1\ \ 1\ \ 0|}=\begin{vmatrix}z_1^1 & z_2^2\\z_2^1 & z_2^2\end{vmatrix},\quad \overline{|0\ \ 1\ \ 1|}=\begin{vmatrix}z_1^2 & z_1^3\\z_2^2 & z_2^3\end{vmatrix},\quad \overline{|1\ \ 0\ \ 1|}=\begin{vmatrix}z_1^1 & z_1^3\\z_2^1 & z_2^3\end{vmatrix} \qquad (7,1)$$

### *7.2 The Gel'fand basis for the invariants of SU(2) is SU(3)*

The generating function of the invariants is:

$$\exp[{}_3\Delta^2_{12}y^2_3 + z^2_3({}_3\Delta^2_{23}x^1_2 + {}_3\Delta^2_{13}y^1_2)] \qquad (7,2)$$

The parameters $y^1_3 = x^1_3 = 0$ are not present in the $\{\phi^{3,[i]}_k(x,y)\}$ of elementary scalars, Therefore the powers of these variables must null:

$$h_{13}-h_{12}=0, \quad h_{12}-h_{23}=0,$$

Then $h_{13} = h_{23} = h_{12}$ and the invariant $H_{(\rho)}$ is the Gel'fand basis:

$$\begin{pmatrix} h_{12} & & h_{12} & & 0 \\ & h_{12} & & h_{22} & \\ & & h_{11} & & \end{pmatrix} \qquad (7,3)$$

***7.3 The expressions of the indices of Gel'fand basis of the invariants***

The expansion of the generating function of 3-j symbols is:

$$H_2(z) = \sum_{(h^i)_2} \begin{pmatrix} h^1_{12} & h^2_{12} & h^3_{12} \\ h^1_{11} & h^2_{11} & h^3_{11} \end{pmatrix} \prod_{i=1}^{3} \frac{(z^{(i)}_1)^{h^i_{12}-h^i_{11}}(z^{(i)}_2)^{h^i_{11}}}{\sqrt{(h^i_{12}-h^i_{11})!(h^i_{11})!}} = N_2 \times ({}_3\Delta^2_{12})^{k_1}({}_3\Delta^2_{13})^{k_2}({}_3\Delta^2_{23})^{k_3} \quad (7,4)$$

With $\qquad k_1 = h_{22}, k_2 = h_{11} - h_{22}, k_3 = h_{12} - h_{11} \qquad (7,5)$

And $\qquad h^i_{12} - h^i_{22} = 2j_i, h^i_{11} - h^i_{22} = j_i - m_i \qquad (7,6)$

We obtain the well known expression of 3-j symbols of Van der Wearden with ρ=1.

# 8. The Gel'fand basis for the invariants of SU(3)

The Gel'fand basis of the invariants polynomials are formed from monomials and function of compatible product of elementary invariant scalars which is reduced to 7 indices variables.

***8.1 The Invariants of the SU(6) Gel'fand basis***

We find for SU (3) seven scalar elementary compatible, which are represented by the following tables:

$$(1) - \overline{\underline{\left|0\ 0\ \right|1\ 0\ \left|1\ 1\right|}} = (z^{(3)}.(z^{(5)} \times z^{(6)}),\ (2) - \overline{\underline{\left|1\ 0\ \right|0\ 0\ \left|1\ 1\right|}} = (z^{(1)}.(z^{(5)} \times z^{(6)})$$

$$(3) - \overline{\underline{\left|1\ 1\ \right|1\ 0\ \left|0\ 0\right|}} = (z^{(3)}.(z^{(1)} \times z^{(2)}),\ (4) - \overline{\underline{\left|1\ 1\ \right|0\ 0\ \left|1\ 0\right|}} = (z^{(5)}.(z^{(1)} \times z^{(2)})$$

$$(5) - \overline{\underline{\left|1\ 0\ \right|1\ 1\ \left|0\ 0\right|}} = (z^{(1)}.(z^{(3)} \times z^{(4)})\quad (6) - \overline{\underline{\left|0\ 0\ \right|1\ 1\ \left|1\ 0\right|}} = (z^{(5)}.(z^{(3)} \times z^{(4)})$$

$$(7) - \overline{\underline{\left|1\ 0\ \right|1\ 0\ \left|1\ 0\right|}} = (z^{(1)}.(z^{(3)} \times z^{(5)}) \qquad (8,1)$$

### *8.2 Generating function of the invariants*

We write the generating function of the invariants in terms of $\phi_k^{n,[i]}(x,y)$ by:

$$
\begin{aligned}
Exp\{\ & x_3^1 x_5^2 x_6^3 y_4^1\ (1) + y_2^1 y_3^1 y_4^1 x_5^2 x_6^3\ (2) + y_4^3 y_5^3 y_6^3\ (3) + y_3^2 y_4^2 x_5^3 y_6^3\ (4) \\
& + y_2^1 x_3^2 x_4^3 y_5^3 y_6^3\ (5) + x_3^1 x_4^2 x_5^3 y_6^3\ (6) + y_2^1 y_4^2 y_6^3 x_3^2 x_5^3\ (7)\ \}
\end{aligned}
\qquad (8,2)
$$

We note the powers of the development of this function $k_1$, $k_2$, ...,$k_6$ and $k_7$.

**A**- The parameters {x, y} that are not present in the elementary scalars $\phi_k^{n,[i]}(x,y)$ must have the power null.

Then the Gel'fand basis for SU(3) invariants is:

$$
\Gamma_6\begin{pmatrix}
h_{13} & h_{13} & h_{13} & 0 & 0 & 0 \\
 & h_{13} & h_{13} & h_{24} & 0 & 0 \\
 & h_{13} & h_{24} & h_{34} & 0 & \\
 & h_{13} & h_{23} & h_{33} & & \\
 & & h_{12} & h_{22} & & \\
 & & h_{12} & & &
\end{pmatrix} =^{s} \Gamma_6\begin{pmatrix} [h]_6 \\ (h)_6 \end{pmatrix} \qquad (8,3)
$$

**B**- The identification of the power of parameters {x, y} of the elementary scalars $\phi_k^{n,[i]}(x,y)$ and the powers in the development of(8,2) give:

$$
\begin{aligned}
& k_1 = (h_{13} - h_{24}) - (h_{12} - h_{23}),\ k_2 = h_{12} - h_{23},\ k_3 = h_{33} \\
& k_4 = h_{22} - h_{33},\ k_5 = h_{34} - h_{33},\ ,\qquad k_6 = h_{24} - h_{23}, \\
& k_7 = (h_{23} - h_{34}) - (h_{22} - h_{33})
\end{aligned}
\qquad (8,4)
$$

**C**-The solution of the system is

$$
\begin{aligned}
& h_{24} = k_3 + k_4 + k_5 + k_6 + k_7, \qquad h_{34} = k_5 + k_3, \\
& h_{13} = k_1 + k_2 + k_3 + k_4 + k_5 + k_6 + k_7, \\
& h_{23} = k_3 + k_4 + k_5 + k_7, \quad h_{33} = k_3, \\
& h_{12} = k_2 + k_3 + k_4 + k_5 + k_7, \quad h_{22} = k_3 + k_4
\end{aligned}
\qquad (8,5)
$$

#### *8.3The homogeneity with respect to the rows and columns*

The invariant of 3j symbolsof SU (3) is given by:

$$
H_{(\rho)} = \sum_{\alpha_i} \begin{pmatrix} \lambda_1\mu_1 & \lambda_2\mu_2 & \lambda_3\mu_3 \\ (\alpha_1) & (\alpha_2) & (\alpha_3) \end{pmatrix}_{(\rho)} V_{(\alpha_1)}^{(\lambda_1\mu_1)}(z^1,z^2) V_{(\alpha_2)}^{(\lambda_2\mu_2)}(z^3,z^4) V_{(\alpha_3)}^{(\lambda_3\mu_3)}(z^{(5)},z^{(6)})^c \qquad (8,6)
$$

#### *A-The homogeneity with respect to the columns($z_{(i)}$,i=1,2,3)*

1- The homogeneity of $V_{(t,tz,y)}^{\lambda\mu}(z^1,z^2)$ give:

$\delta_1=t_z+(Y/3)+(\lambda+2\mu)/3$ , $\delta_2= -t_z+(Y/3)+(\lambda+2\mu)/3$, $\delta_3=-(Y/3)+(\lambda+2\mu)/3$ (8,7)

2- The homogeneity in equation (8,6) give:

$$P = (t_{1z} + t_{2z} - t_{3z}) + (Y_1 + Y_2 - Y_3)/3 + P1,$$
$$P = -(t_{1z} + t_{2z} - t_{3z}) + (Y_1 + Y_2 - Y_3)/3 + P1$$
$$P = -(Y_1 + Y_2 - Y_3)/3 + P1$$

With $P = k_1 + k_2 + k_3 + k_4 + k_5 + k_7$ ,

We deduce that: $(t_{1z} + t_{2z} - t_{3z}) = (Y_1 + Y_2 - Y_3) = 0$

And $P = P1 = (\lambda_1 + \lambda_2 + \mu_3 + 2(\mu_1 + \mu_2 + \lambda_3))/3$ (8,8)

***B-The homogeneitywith respect to the rows ($z^{(i)}$,i=1,2,..,6)***

1- The homogeneityof $H_\rho$ give:

$$k_1+k_2=\lambda_3,\ k_1+k_4+k_7=\lambda_2,\quad k_2+k_5+k_7=\lambda_1$$
$$k_3+k_6+k_7=\mu_3,\ k_5+k_6=\mu_2,\quad k_3+k_4=\mu_1 \qquad (8,9)$$

The addition of these equations is:

$$2P+k_7= \lambda_1+\lambda_2+\lambda_3+\mu_1+\mu_2+\mu_3$$

2- We find

$$k_7 = (\lambda_1 + \lambda_2 + \mu_3 - (\mu_1 + \mu_2 + \lambda_3))/3$$

And $k_2=\lambda_3-k_1$, $k_3=\mu_1- k_4$, $k_4=\lambda_2- k_7-k_1$,
$k_5= \lambda_1- k_2- k_7$, $k_6=\mu_2-(\lambda_1-k_7-k_2)$; (8,10)

***8.4 Expression of the invariants in Gel'fand basis***

$$h_{24} = 2\lambda_3 + \mu_2 + \lambda_2 - \lambda_1 + k_7 - 2k_1,\qquad h_{34} = \lambda_3,$$
$$h_{13} = P,\qquad h_{23} = \lambda_3 + \lambda_2 - k_1,\qquad h_{33} = \lambda_3 - k_1,$$
$$h_{12} = \lambda_3 + \mu_1 + k_7,\qquad h_{22} = \lambda_3 + \lambda_1 - k_7 - 2k_1 \qquad (8,11)$$

## 9. The Gel'fand basis of SU(n) invariants and It's number formula

The generalization of the basis of SU (2) and SU(3) to SU(n) requires that Δ(z) is only a function of $z^{(i)} = (z_1^i, z_2^i, ..., z_n^i),\ i = 1,..,n-1,\ z_i^j \in C$ and the invariant of SU (n) is expressed in terms of binary numbers of 3(n-1) boxes.

***9.1 The Invariants of SU(4)***

$$1)\ \overline{|100|111|000|},\ 2)\ \overline{|110|110|000|},\ 3)\ \overline{|111|100|000|},\ 4)\ \overline{|100|000|111|}$$
$$5)\ \overline{|110|000|110|},\ 6)\ \overline{|111|000|100|},\ 7)\ \overline{|000|100|111|},\ 8)\ \overline{|000|110|110|}$$
$$9)\ \overline{|000|111|100|},\ 10)\ \overline{|100|100|110|},\ 11)\ \overline{|100|110|100|},\ 12)\ \overline{|110|100|100|}$$

### *9.2 The parameters of the elementary scalars*

The parameters $\{x, y\}$ of the elementary scalars $\phi_k^{n,[i]}(x, y)$ are:

1) $y_2^1 y_3^1 y_7^4 y_8^4 y_9^4 x_4^2 x_5^2 x_6^4$ 2) $x_4^3 x_5^4 y_3^2 y_6^4 y_7^4 y_8^4 y_9^4$, 3) $y_5^4 y_6^4 y_7^4 y_8^4 y_9^4$, 4) $y_2^1 y_3^1 y_4^1 y_5^1 y_6^1 x_7^2 x_8^3$
5) $x_7^3 x_8^4 y_3^2 y_4^2 y_5^2 y_6^2 y_9^4$ 6) $x_7^4 y_4^3 y_5^3 y_6^3 y_8^4 y_9^4$ 7) $x_4^1 x_7^2 x_8^3 x_9^4 y_5^1 y_6^1$, 8) $x_4^1 x_5^2 x_7^3 x_8^3 y_6^2 y_9^4$
9) $x_4^1 x_5^2 x_6^3 x_7^4 y_8^4 y_9^4$, 10) $x_4^2 x_7^3 x_8^3 y_2^1 y_3^1 y_5^2 y_6^2 y_9^4$, 11) $x_4^2 x_5^3 x_7^4 y_2^1 y_3^1 y_6^3 y_8^4 y_9^4$, 12) $x_4^3 x_7^4 y_3^2 y_5^3 y_6^3 y_8^4 y_9^4$

### *9.3 The Gel'fand basis of the Invariants of SU(4)*

The Gel'fand basis of the invariants of SU(4) in terms of SU(9) basis is:

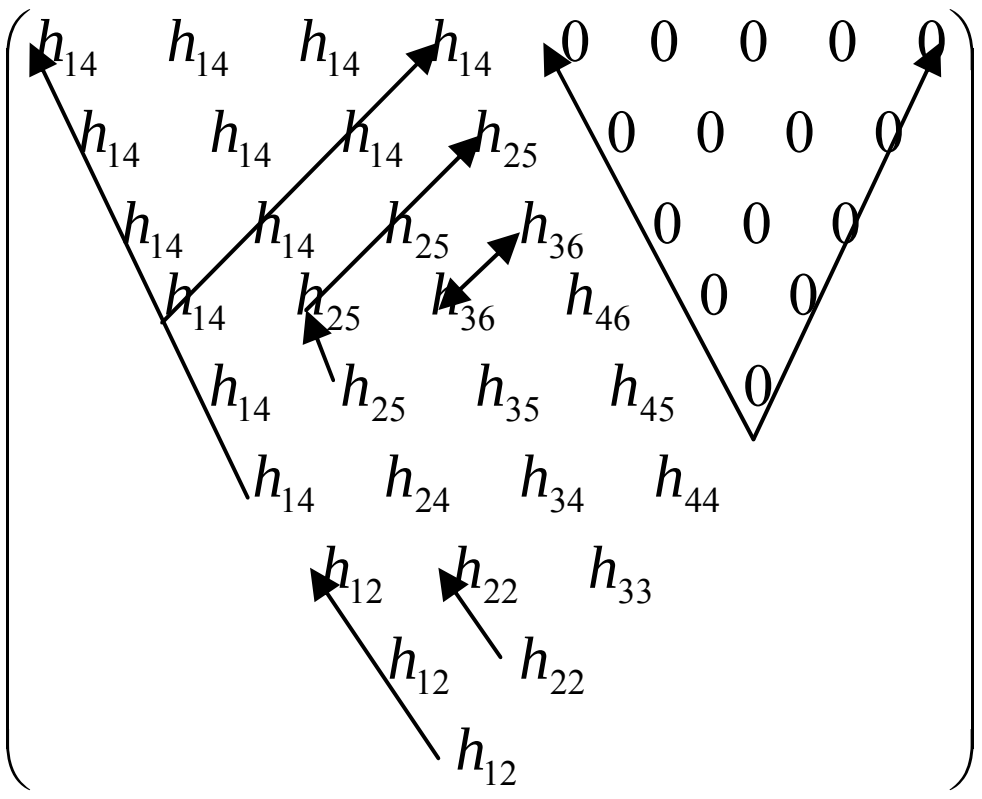

(9,1)

### *9.4 The Gel'fand basis for the invariants of SU(n)*

After we have made all the calculations also for SU (5) and SU (6) we found that we can write the basis of Gel'fand invariants as:

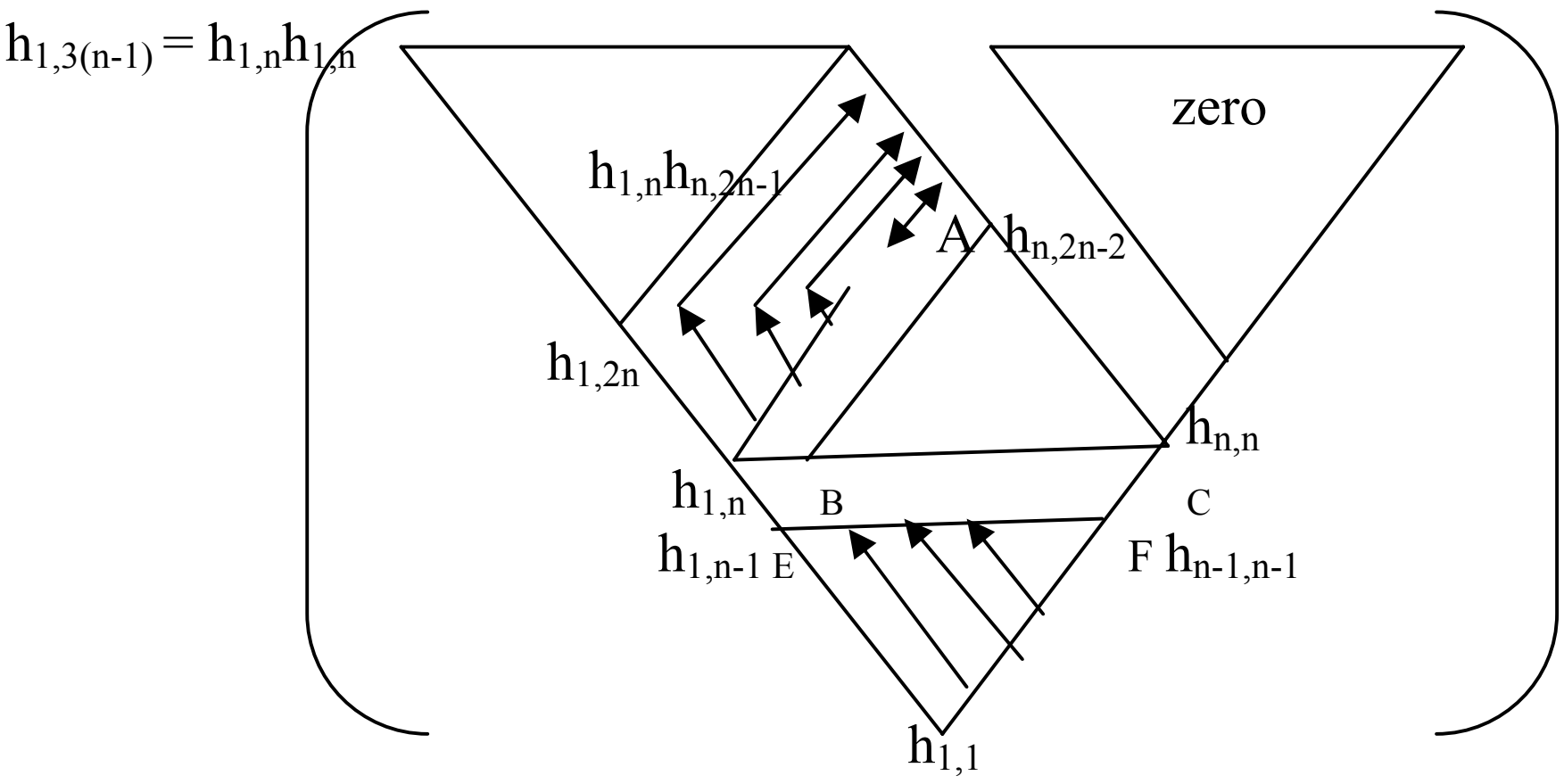

(9,2)

It is interesting to note that we obtain a form of double Gel'fand basis similar to the tensor method of Biedenharn [24].

***9.5 The formula of the number of invariants***
The number of indices is:
a- On the straight line$h_{1n}h_{n,2n-1}$ there are (n-1) indices because $h_{n,2n-1}=h_{n-1,2n-2}$
b- On the straight line EF there are (n-1) indices
c-In the triangle ABC there are n(n-1) / 2 indices
The total sum of the indices is:

$$N= n(n-1)/2+2(n-1)=(n-1)(n+4)/2. \qquad (9,3)$$

We find for N =2,3,4,5,6 the following values: 3,7,12,18,25.
We obtain by a simple Maple program of (6,8) the same numerical values of the formula (9,3) whatever N.

## 10. The isoscalar of $(\lambda_1 0, \lambda_2\mu_2; \lambda_3\mu_3)$ symbols

To overcome the difficulties of the methods already proposed [9, 12,17-18] we propose a simple idea which assumes the independence of variables $\{\Delta_n^i(z)\}, \ i=1,2,3$
in the expression(6,4)for the calculation of Wigner symbols of SU (n).
Our proposal is to replace ${}^s\Delta_n^i(z)$ in(6.4)by the elementary invariants of the space of parameters ${}^s\phi_n^i$.
The symbols $(\lambda_1 0, \lambda_2\mu_2; \lambda_3\mu_3)$ was calculate by Moshinsky [19] using the basis of the oscillator. Then Resnikoff [17],following the well known Bargmann work on SU(2),transpose Moshinsky work on SU (3) to Fock-Bargmann space.
The result of calculation by our method gives the expression of Wigner's symbols and the isoscalar factors with only two summations.
It is important to note that the calculations of the symbols $(\lambda_1 0, \lambda_2 0; \lambda_3\mu_3)$ is done by a direct calculation using the Gauss integral [22-23].

***10.1 The Van der Wearden invariants of SU(2) in the space of parameters***
We replace ${}^s\phi_n^i$(x, y) by ${}^s\phi_2^i(u^1,u^2)$ in the expression (7,4) and the elementary invariants in the space of parameters are:

$$\overline{\underline{\begin{vmatrix}1 & 1 & 0\end{vmatrix}}} \Rightarrow \begin{vmatrix} z_1^1 & z_2^2 \\ z_2^1 & z_2^2 \end{vmatrix} = \Delta_1^1\Delta_2^2 - \Delta_2^1\Delta_1^2 \Rightarrow \Xi(1,2) = \begin{vmatrix} y1(2,1) & y2(2,1) \\ x1(2,1) & x2(2,1) \end{vmatrix} \qquad (10,1)$$

And
$$\Xi(i,j) = [u^i u^j] = \begin{vmatrix} yi(2,1) & yj(2,1) \\ xi(2,1) & xj(2,1) \end{vmatrix} = \begin{vmatrix} u_1^i & u_1^j \\ u_2^i & u_2^j \end{vmatrix}$$

The Van der Wearden invariants is:

$$H_2(t) = \sum_{m_i}\left[\prod_{i=1}^3 \varphi_{t^i,t_0^i}(u^i)\right]\begin{pmatrix} t^1 & t^2 & t^3 \\ t_0^1 & t_0^2 & t_0^3 \end{pmatrix} = \frac{[u^2u^3]^{(T-2t^1)}[u^3u^1]^{(T-2t^2)}[u^1u^2]^{(T-2t^3)}}{N_2} \qquad (10,2)$$

With $N_2 = \sqrt{(T+1)!(T-2t^1)!(T-2t^2)!(T-2t^3)}$, $T = t^1 + t^2 + t^3$

The invariance by permutations of (3-t)symbols imposes the condition $m_1+m_2+m_3 = 0$.

***10.2 The generating function of 3-j symbols and the isoscalar of SU(3)***

The expression (8,6) is invariant by permutation of columns if:

$$\boxed{\sum_{(h^i)_3} \begin{pmatrix} [h^1]_3 & [h^2]_3 & [h^3]_3 \\ (h^1)_3 & (h^2)_3 & (h^3)_3 \end{pmatrix}_\rho \prod_{i=1}^{3} \Gamma_3 \begin{pmatrix} [h^i]_3 \\ (h^i)_3 \end{pmatrix} \left( {}^{S}\phi^i \right) = N_3 \prod_{i=1}^{7} [W^i]^{k_i}} \quad (10,3)$$

The Wigner coefficients of SU (3) is:

$$\begin{pmatrix} (\lambda_1\mu_1) & (\lambda_2\mu_2) & (\lambda_3\mu_3) \\ (\alpha_1) & (\alpha_2) & (\alpha_3) \end{pmatrix}_\rho = \begin{Bmatrix} (\lambda_1\mu_1) & (\lambda_2\mu_2) & (\lambda_3\mu_3) \\ [t_1, y_1] & [t_2, y_2] & [t_3, y_3] \end{Bmatrix}_\rho \begin{pmatrix} t_1 & t_2 & t_3 \\ t_{1z} & t_{2z} & t_{3z} \end{pmatrix} \quad (10,4)$$

In this case we obtain the isoscalar factor {3(λμ, t, y)} of SU (3) but we must change the result of the calculations by $\lambda_3 \leftrightarrow \mu_3$, $(t, t_z, y) \to (t, -t_z, -y)$.

The calculations of $N_g$ is very long [17], but we give only the result:

$$N_g = \left[ \frac{(P+2)!(k_1+k_2+k_7)!(k_1+k_6+k_7+1)!(k_5+k_6+k_7+1)!}{2k_7!k_1!k_2!k_5!k_6!(k_1+k_7+1)!(k_6+k_7+1)!} \right]^{1/2} \quad (10,5)$$

From the expression (8,9) we deduce $k_3=k_4= 0$ if $\mu_1=0$.

We seek to determine the first term first and then the second term of(10,3).

***10.3 Development of the first member of the generating function of SU(3)***

***10.3.1 Calculation of*** $V^{\lambda\mu}_{(t,tz,y)}(\phi^1, \phi^2)$

$$V^{\lambda\mu}_{(t,tz,y)}(\phi^1, \phi^2) = N(\lambda\mu;\alpha)(-1)^s \times \sum_k \begin{pmatrix} m \\ k \end{pmatrix} \frac{(\mu - s)! r!}{(\mu - s - k)![r-(n-k)]!}$$

$$\times \prod_{\lambda=2}^{3} \prod_{\mu=1}^{\lambda-1} ((x^\mu_\lambda)^{h_{\mu\,\lambda} - h_{\mu\,\lambda-1}} (y^\mu_\lambda)^{h_{\mu\,\lambda-1} - h_{\mu+1\,\lambda}}) \quad (10,6)$$

Using the known expression of hypergeometric functions F(α,β; γ; 1) we write:

$$V^{\lambda\mu}_{(t,tz,y)}(\phi^1, \phi^2) = N(\lambda\mu;\alpha)(-1)^s \frac{(r+\mu-s)!}{(r-m+\mu-s)!} \times \prod_{\lambda=2}^{3} \prod_{\mu=1}^{\lambda-1} ((x^\mu_\lambda)^{h_{\mu\,\lambda} - h_{\mu\,\lambda-1}} (y^\mu_\lambda)^{h_{\mu\,\lambda-1} - h_{\mu+1\,\lambda}}) \quad (10,7)$$

using also (2,8) and $2t = \mu + (r - s)$ we obtain:

$$V^{\lambda\mu}_{(t,tz,y)}(\phi^1, \phi^2) = N_v(\lambda\mu; t, y) \prod_{\lambda=2}^{3} \prod_{\mu=1}^{\lambda-1} ((x^\mu_\lambda)^{h_{\mu\,\lambda} - h_{\mu\,\lambda-1}} (y^\mu_\lambda)^{h_{\mu\,\lambda-1} - h_{\mu+1\,\lambda}}) \times \frac{1}{\sqrt{(t+t_0)!(t-t_0)!}} \quad (10,8)$$

With $N_v(\lambda\mu;t,y)=\left\{\frac{(\lambda+1)!(2t+1)!}{r!s!(\mu-s)!(\lambda-r)!(\mu+r+1)!(\lambda+\mu-s+1)!}\right\}^{\frac{1}{2}}(2t)!$ (10,9)

***10.3.2 Expression of the First member***

using:

$$\prod_{\lambda=2}^{3}\prod_{\mu=1}^{\lambda-1}(x_\lambda^\mu)^{h_{\mu\,\lambda}-h_{\mu\,\lambda-1}}(y_\lambda^\mu)^{h_{\mu\,\lambda-1}-h_{\mu+1\,\lambda}})=\prod_{\mu=1}^{2}((x_i^\mu)^{h^i_{\mu\,3}-h^i_{\mu\,2}}(y_i^\mu)^{h^i_{\mu\,2}-h^2_{\mu+1\,2}})\times((x_i^1)^{h^i_{1\,2}-h^i_{1,1}}(y_i^1)^{h^i_{1,1}-h^i_{2,2}})$$

The first expression of (10.3) may be written :

$$\sum\{3(\lambda\mu;t,t_0,y)\}\prod_{i=1}^{3}\Gamma_3\begin{pmatrix}[h^i]_3\\(h^i)_3\end{pmatrix}\left({}^s\phi^i\right)=\sum\{3(\lambda\mu;t,y)\}\{\prod_{i=1}^{3}\{N_v(\lambda_i\mu_i;t_i,y_i)\}\times$$

$$\left.\prod_{\mu=1}^{2}((x_i^\mu)^{h^i_{\mu\,3}-h^i_{\mu\,2}}(y_i^\mu)^{h^i_{\mu\,2}-h^2_{\mu+1\,2}}\right\}\times\sum(3-t)\prod_{i=1}^{3}((x_i^1)^{h^i_{\mu\,3}-h^i_{\mu\,2}}(y_i^1)^{h^i_{\mu\,2}-h^2_{\mu+1\,2}} \quad (10,10)$$

With the help of (10.2) we obtain:

$$\sum\{3(\lambda\mu;t,t_0,y)\}\prod_{i=1}^{3}\Gamma_3\begin{pmatrix}[h^i]_3\\(h^i)_3\end{pmatrix}\left({}^s\phi^i\right)=\sum\{\left[\{3(\lambda\mu;t,y)\}\times\prod_{i=1}^{3}[N_v(\lambda_i\mu_i;t_i,y_i)]\times\right.$$

$$\left.\prod_{\mu=1}^{2}((x_i^\mu)^{h^i_{\mu\,3}-h^i_{\mu\,2}}(y_i^\mu)^{h^i_{\mu\,2}-h^2_{\mu+1\,2}}\right]\times[H_2(t)]\} \quad (10,11)$$

***10.4 Expression of the second member of the generating function of SU(3)***

***10.4.1 Calculus of the invariants in the space of parameters*** ${}^s\phi_6^i$

To determine the images of invariants [W] in the space of parameters we write:

$$\overline{\underline{|1\ \ 0\ \ 1\ \ 1\ \ 0\ \ 0|}}(z)=\begin{vmatrix}z_1^1 & z_1^3 & z_1^4\\ z_2^1 & z_2^3 & z_2^4\\ z_3^1 & z_3^3 & z_3^4\end{vmatrix}=\Delta_1^1\Delta_{23}^{34}-\Delta_2^1\Delta_{13}^{34}+\Delta_3^1\Delta_{12}^{34})\,(10,1) \quad (10,12)$$

$$\Rightarrow W_1=y1(3,1)x2(3,2)\Xi(1,2)+x1(3,1)y2(3,2)$$

We apply the same method for the calculation of the image of the invariants and for simplicity we change $yi(3,m), xi(3,m)$ *by* $y_m^i$ *and* $x_m^i$.

$$(1)\Rightarrow W_1=y_1^2x_2^3[u^2u^3]+x_1^2y_2^3\ \ (2)\Rightarrow W_2=y_1^1x_2^3[u^1u^3]+x_1^1y_2^3$$

$$(3)\Rightarrow W_3=-y_1^2x_2^1[u^1u^2]+x_1^2y_2^1\ \ (4)\Rightarrow W_5=-y_1^3x_2^1[u^1u^3]+x_1^3y_2^1$$

$$(5)\Rightarrow W_3=y_1^1x_2^2[u^1u^2]+x_1^1y_2^2\ \ (6)\Rightarrow W_6=-y_1^3x_2^2[u^2u^3]+x_1^3y_2^2$$

$$(7)\Rightarrow W_7=x_1^3y_1^1y_1^2[u^1u^2]-x_1^2y_1^1y_1^3[u^1u^3]+x_1^1y_1^2y_1^3[u^2u^3] \quad (10,13)$$

### *10.4.2 Expression of the second member of (10,3) with $\mu_1=0$*

We put :

$$[W]=\frac{(W_1)^{k_1}((W_2)^{k_2}(W_5)^{k_5}(W_6)^{k_6}(W_7)^{k_7}}{k_1!k_2!k_5!k_6!k_7!} \qquad (10,14)$$

We note that the normalization of [W] can be done by several methods [9,17,23]. And after the development of [W] we find:

$$\begin{array}{ll}
y_1^1 \Rightarrow i+j+m+n=h_{12}^1-h_{23}^1, & y_2^1 \Rightarrow h_{22}^1-h_{33}^1=0 \\
y_1^2 \Rightarrow k+k_7-n=h_{12}^2-h_{23}^2, & y_2^2 \Rightarrow k_5-i+k_6-l=h_{22}^2-h_{33}^2, \\
y_1^3 \Rightarrow l+k_7-m=h_{12}^3-h_{23}^3, & y_2^3 \Rightarrow k_2-j+k_1-k=h_{22}^3-h_{33}^3 \\
x_1^1 \Rightarrow k_1-i+k_2-j+k_7-m-n=h_{13}^1-h_{12}^1, & x_2^1 \Rightarrow h_{23}^1-h_{22}^1=0 \\
x_1^2 \Rightarrow k_5-k+n=h_{13}^2-h_{12}^2, & x_2^2 \Rightarrow i+l=h_{23}^2-h_{22}^2, \\
x_1^3 \Rightarrow k_6-l+m=h_{13}^3-h_{12}^3, & x_2^3 \Rightarrow k+j=h_{23}^3-h_{22}^3
\end{array} \qquad (10,15)$$

Then we find:

$$\begin{array}{c}
k_1+k_2=\lambda_3\,,\quad k_5+k_6=\mu_2\,, \\
k_6+k_7=\mu_3,\quad k_1+k_7=\lambda_2,\quad k_2+k_5+k_7=\lambda_1
\end{array} \qquad (10,16)$$

We get the same intermediate expressions of reference (17) which proves the validity and importance of our proposal.

### *10.5 Expression of the isoscalar of $(\lambda_1 0, \lambda_2\mu_2, \lambda_3\mu_3)$*

The development of [W] is:

$$[W]=(-1)^T\sum_{i,j}\frac{(-1)^{l+n}}{i!(k_5-i)!\,j!(k_2-j)!k!(k_1-k)!l!(k_6-l)!m!n!(k_7-m-n)!}\times$$

$$\left\{\left[\prod_{\lambda=1}^{2}\prod_{i=1}^{3}(x_i^{\mu})^{h_{\mu\,3}^i-h_{\mu\,2}^i}(y_i^{\mu})^{h_{\mu\,2}^i-h_{\mu+1\,2}^2}\right]\left[N_2\times H_2(t)\right]\right\} \qquad (10,17)$$

Comparing this expression with (10.6) we find the isoscalar factors of SU (3):

$$\left\{\begin{matrix}(\lambda_1\mu_1) & (\lambda_2\mu_2) & (\lambda_3\mu_3)\\ [t_1,y_1] & [t_2,y_2] & [t_3,y_3]\end{matrix}\right\}=(-1)^T\frac{N_2}{\prod_{i=1}^{3}\left[N_v(\lambda_i\mu_i;t_i,y_i)\right]}\times$$

$$\left[\sum_{i,j}\frac{(-1)^{l+n}}{i!(k_5-i)!\,j!(k_2-j)!k!(k_1-k)!l!(k_6-l)!m!n!(k_7-m-n)!}\right] \qquad (10,18)$$

With

$$\begin{array}{ll}
l=h_{23}^2-h_{22}^2-i, & m=h_{13}^3-h_{12}^3+h_{23}^2-h_{22}^2-i-k_6 \\
k=h_{23}^3-h_{22}^3-j, & n=h_{13}^2-h_{12}^2+h_{23}^3-h_{22}^3-j+k_1
\end{array} \qquad (10,19)$$

We get not only Wigner symbols but more the isoscalar factors of SU (n) with only two summations.

Finally, it is important to note that the generating function method [23] that we develop has a variety of applications [22-23] and in particular the unitary group theory and it is clear that this method requires only the undergraduate level.

## References

[1] J.P. Elliott, Proc, Roy. Soc. (London)A245,128,562 (1958)
[2] A. Rougé, "Introduction à la physique subatomique", Ed. Ellipse (1995).
[3] W. Greiner, "Quantum Mechanics (Symmetries) "Ed. Springer (1994)
[4] I.M. Gel'fand and M. L. Zeitlin, Dokl.Akad.Nauk 71, 825 (1950).
[5] A. O. Barut and R. Raczka, Theory of group representations and applications PWN- Warszawa (1980).
[6] G. E. Baird and L. C. Biedenharn, J. Math. Phys., 4(1963) 1449; 5(1964) 1723;
[7] M. Hage-Hassan, "A note on Quarks and numbers theory " arXiv: Math-Physics / 1302.6342 (2013).
[8] M. Hage-Hassan, J. Phys. A, 12 (1979)1633, J. Phys. A, 16 (1983)1835
[9] W.J. Holman III and L.C. Biedenharn, "Group Theory and its Applications"(1971), Ed. Loebl (New York: Academic Press)
[10] K. T. Hecht and L. C. Biedenharn, J. Math. Phys., 31 (1990) 2781
[11] Feng Pan† and J. P. Draayer , arXiv:quant-ph/9704014 v1 8 Apr 1997
[12] D.J. Rowe and C. Bahri, J. Phys. A, 41, 9 (2000)6544
[13] D.J. Rowe, B.C. Sanders, H. de Guise arXiv: math-phy/9811012v2 (1996, 2006)
[14] C. Bahri, D.J. Rowe, J. P. Draayer, Computer Physics Communications 159(2004)121
[15] J. Schwinger, In "Quantum Theory of angular momentum" Ed. L.C. Biedenharn and H. Van Dam, Acad. Press, NY, (1965)
[16] V. Bargmann, "On the representations of the Rotation group" Rev. Mod. Phys.34, 4, (1962)
[17] M. Resnikoff, J. math. Phys. 8(1967)63
[18] C. K. Chew and R. T. Sharp, Can. J. Phys., 44 (1966) 2789
[19] M. Moshinsky, J. Math. Phys., 4 (1963) 1128; Rev. Mod. Phys., 34 (1962) 813
[20] J. Nagel and M. Moshinsky, J. math. Phys. 6(1965)682
[21] C.J. Henrich, J. Math. Phys. 16, 2271 (1975).
[22] M. Hage-Hassan, "On Fock-Bargmann space, Dirac delta function, Feynman propagator, angular momentum and SU(3) multiplicity free" arXiv: [math-ph] /0909.0129, (2010),
[23] M. Hage-Hassan, « Generating function method and its applications to Quantum, Nuclear and the Classical Groups " arXiv: Math-Physics /1203.2892 (2012).
[24] D. Lurie and A. J. Macfarlane, J. Math. Phys. 5, 565 (1964)